\documentclass[12pt]{iopart}
\usepackage{iopams,epsfig,wrapfig}

\usepackage{graphicx}
\usepackage{latexsym}
\usepackage{bm}

\usepackage{color}

\begin{document}

\title{CGC, Hydrodynamics, and the Parton Energy Loss}
\author{Tetsufumi Hirano\dag\ and Yasushi Nara\ddag}

\address{\dag\ RIKEN BNL Research Center,
    BNL,
Upton, New York 11973, USA}
\address{\ddag\ Department of Physics, University of Arizona,
 Tucson, Arizona 85721, USA}

\begin{abstract}
Hadron spectra in Au+Au collisions at RHIC
are calculated by hydrodynamics with 
initial conditions
from the Color Glass Condensate (CGC).
Minijet components with parton energy loss in medium
are also taken into account by using 
parton density obtained from hydrodynamical simulations.
We found that
CGC provides a good initial condition
for hydrodynamics in Au+Au collisions at RHIC.
\end{abstract}

High energy heavy ion collisions involve different aspects
according to the relevant energy or time scale.
There already exist many theoretical approaches to understand
numerous RHIC data.
In this work, we consider in particular
physics of gluon saturation, hydrodynamic evolution,
and the energy loss of hard partons in the medium.
Our goals are to combine them and to take a step to a unified
understanding of
the dynamical aspect of high energy heavy ion collisions.

At soft region $p_T < 2$ GeV/c, where
 bulk dynamics is governed,
hydrodynamical description~\cite{Huovinen:2003fa}
is successful in describing
the elliptic flow
at low $p_T$, up to semi-central collisions,
 and mid-rapidity~\cite{Hirano:2001eu} at RHIC.
This is one of the strongest indications of an early
thermalization
of the quark gluon plasma (QGP) at RHIC.
Hydrodynamics also predicts that the scaled elliptic flow,
which is defined as the second harmonics $v_2$ divided by
initial spatial eccentricity $\varepsilon$,
becomes almost constant around 0.2~\cite{Kolb:2000sd}.
 The experimental data reaches the hydrodynamic limit
for the first time
in central and semi-central collisions at RHIC energies~\cite{STAR:ecc}.
On the other hand,
minijets go across the expanding matter
and lose their energies 
(jet quenching) in heavy ion collisions~\cite{GVWZKW}.
Observed large suppression of hadron spectra~\cite{phenix:pi0}
and disappearance of the away-side peak in azimuthal correlation
functions at mid-rapidity~\cite{STAR:btob}
have been interpreted
as a consequence of parton energy loss in the medium.
Indeed, 
Cronin enhancement of the hadron spectra and 
existence of the back-to-back
correlation at mid-rapidity in $dA$ collisions at RHIC~\cite{rhic:dA}
support the importance of the strong final state interaction
in $AA$ collisions.
The current RHIC data strongly
suggest that the initial parton density is large.
What is an origin of the large density in Au+Au collisions at RHIC?
The bulk particle production is dominated by the small $x$
modes in the nuclear wave function, where $x$ is a momentum
fraction of the incident particles.
It is well known that gluon density increases rapidly 
with decreasing $x$ by the BFKL cascade
until gluons begin to overlap in phase space
where nonlinear interaction becomes important~\cite{GLR83}.
These gluons
form the Colour Glass Condensate (CGC)~\cite{MV}.
%
Remarkably, the CGC results on the global observables such as
the centrality, rapidity and the energy dependences of charged
hadron multiplicities agree with the RHIC data~\cite{KLN}.
It has been shown that
the classical wave function in the MV model contains
Cronin enhancement
and the quantum evolution in $x$ makes
 the spectrum suppressed~\cite{croninCGC}.

From the above considerations,
the CGC, hydrodynamics,
and the energy loss of hard partons are key ingredients to describe
the RHIC physics
and must be closely related with each other.
For example, the CGC could be a good initial
condition for thermalization
because it produces a large number of gluons.
Thus these gluons are
responsible to the large suppression
of jet spectra.
In this work we assume that
the origin of thermalized 
partonic matter is the CGC in high energy heavy ion collisions,
and use it as an initial condition in the hydro+jet model~\cite{HiranoNara}.
With this approach we expect to get deeper understanding of the
dynamical aspect of the heavy ion collisions.
In addition, some of the problems which are inherent in a particular
approach can be removed.
We employ the $k_T$ factorized formula
 along the line of Kharzeev, Levin, and Nardi (KLN)~\cite{KLN}
for the computation of the gluon rapidity distribution which
is given by
\begin{equation}
\frac{dN_g}{d^2x_{\perp}dy} = 
   \frac{4\pi^2N_c}{N_c^2-1} \int\frac{d^2p_T}{p^2_T}
   \int d^2k_T \alpha_s
   \phi_A(x_1,k_T^2)\phi_B(x_2,(p_T-k_T)^2),
 \label{eq:ktfac}
\end{equation}
where
$x_{1,2}=p_T\exp(\pm y)/\sqrt{s}$ with $y$ and
$p_T$ are a rapidity
and a transverse momentum of a produced gluon.
We assume that the system of initially produced gluons
reaches local thermailzed state at a short time scale.
Although the produced gluons
will reach the thermalized state through the dissipative
processes in the realistic situations,
the description of non-equilibrium phenomena
is beyond the scope of the present paper.
We also assume that 
the shape of the rapidity density distribution
is not changed during the system is thermalized.
Therefore, we take initial conditions
from gluon distribution
obtained from Eq.~(\ref{eq:ktfac}).
Assuming Bjorken's ansatz $y=\eta_{\mathrm{s}}$ where $\eta_{\mathrm{s}}$
is the space-time rapidity,
we obtain the number or the energy density
for gluons at each space-time point.

The initial transverse energy per particle yields
$E_T/N_g \sim 1.6$ GeV at $y=0$ from Eq.~(\ref{eq:ktfac}).
This is within a range estimated
in a numerical simulation of
the classical Yang-Mills equation~\cite{KNV}.
It should be noted that the assumption of the thermalization
in CGC gluon distribution is to reduce
the transverse energy per particle from
$E_T/N_g=1.6$ to $E_T/N_g \approx 1$ within our present parameters.
The effect of the hydrodynamic afterburner is to reduce
the transverse energy per particles
due to $pdV$ work and yields $(dE_T/dy)/(dN/dy)|_{y=0} = 0.54$ GeV.
The rapidity distribution becomes slightly wider.
Our result supports that
KLN calculation~\cite{KLN} which is based on the assumption of parton-hadron
duality is a good approximation on the rapidity distributions.
%
%
\begin{figure}[ht]
\includegraphics[width=3.0in]{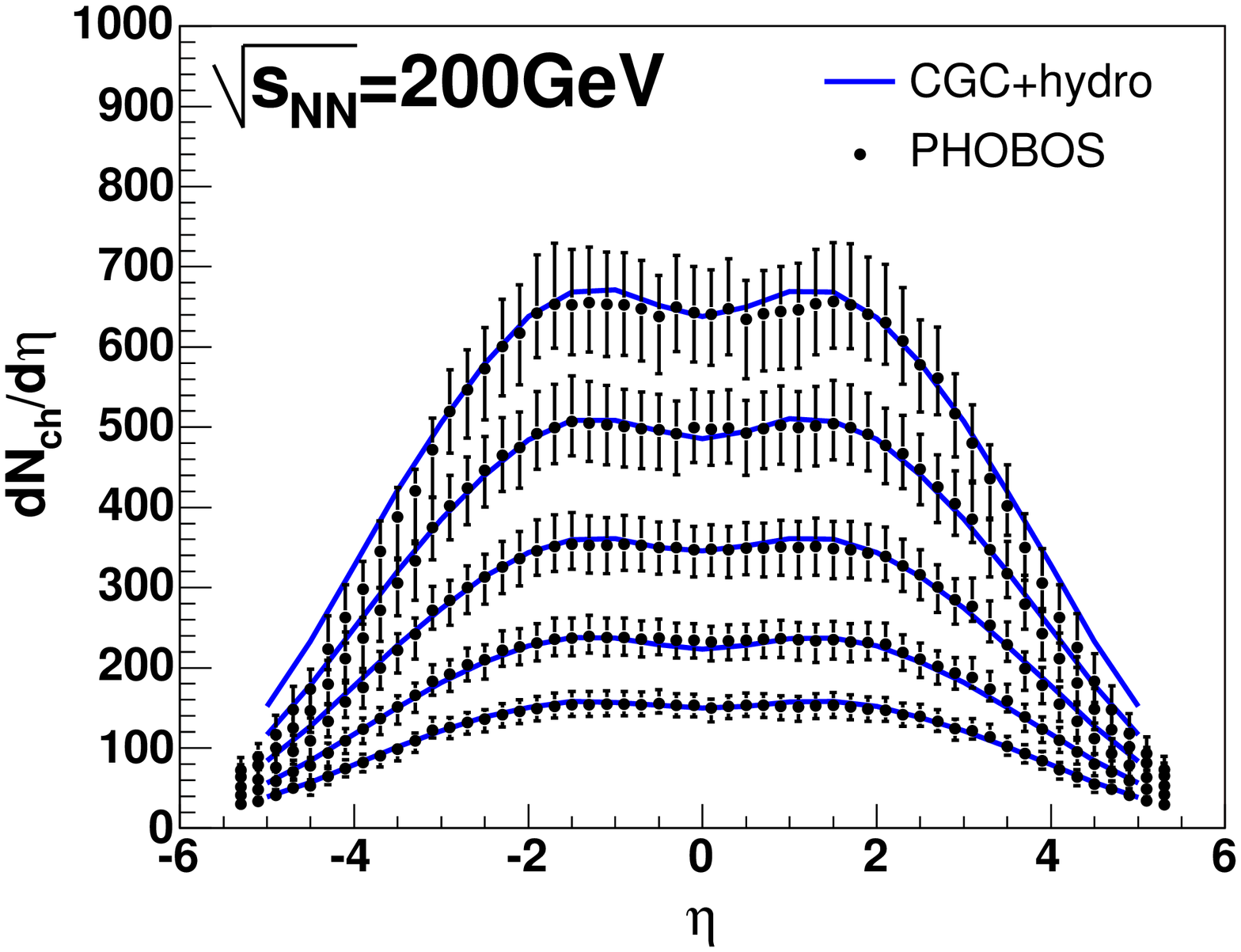}
\includegraphics[width=3.0in]{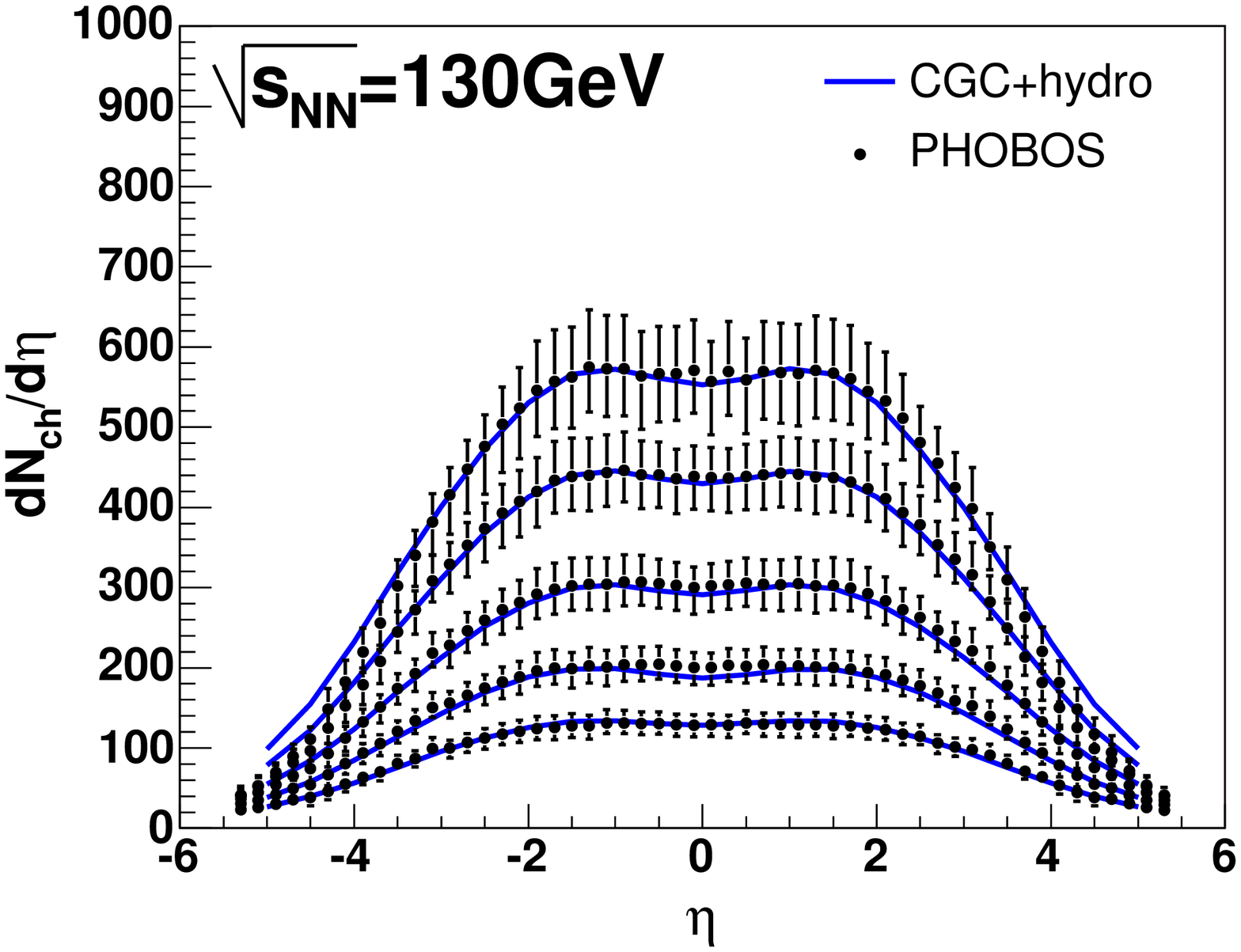}
\caption{
Pseudorapidity distributions of charged hadrons
 in Au + Au  collisions at $\sqrt{s_{NN}}=130$ and 200 GeV
is compared to the PHOBOS data~\cite{PHOBOS:dndeta}.
Impact parameters which correspond to $\langle N_{\mathrm{part}}\rangle$
from PHOBOS are used in the calculations.
}
\label{fig:dndeta200}
\end{figure}
In Fig.~\ref{fig:dndeta200}, pseudorapidity distributions
of charged hadrons in Au + Au collisions at both $\sqrt{s_{NN}}=130$
and 200 GeV are compared with the PHOBOS data~\cite{PHOBOS:dndeta}.
$k_T$ factorization approach in the CGC provides
very good initial conditions
for the hydrodynamical simulations which reproduce rapidity,
centrality and energy dependences of multiplicity.
It should be emphasized that it is very hard to find such
a good initial condition which fits the data 
with the same quality
as the CGC one presented here.
                                                                                
\begin{figure}[t]
\includegraphics[width=3.0in,height=2.3in]{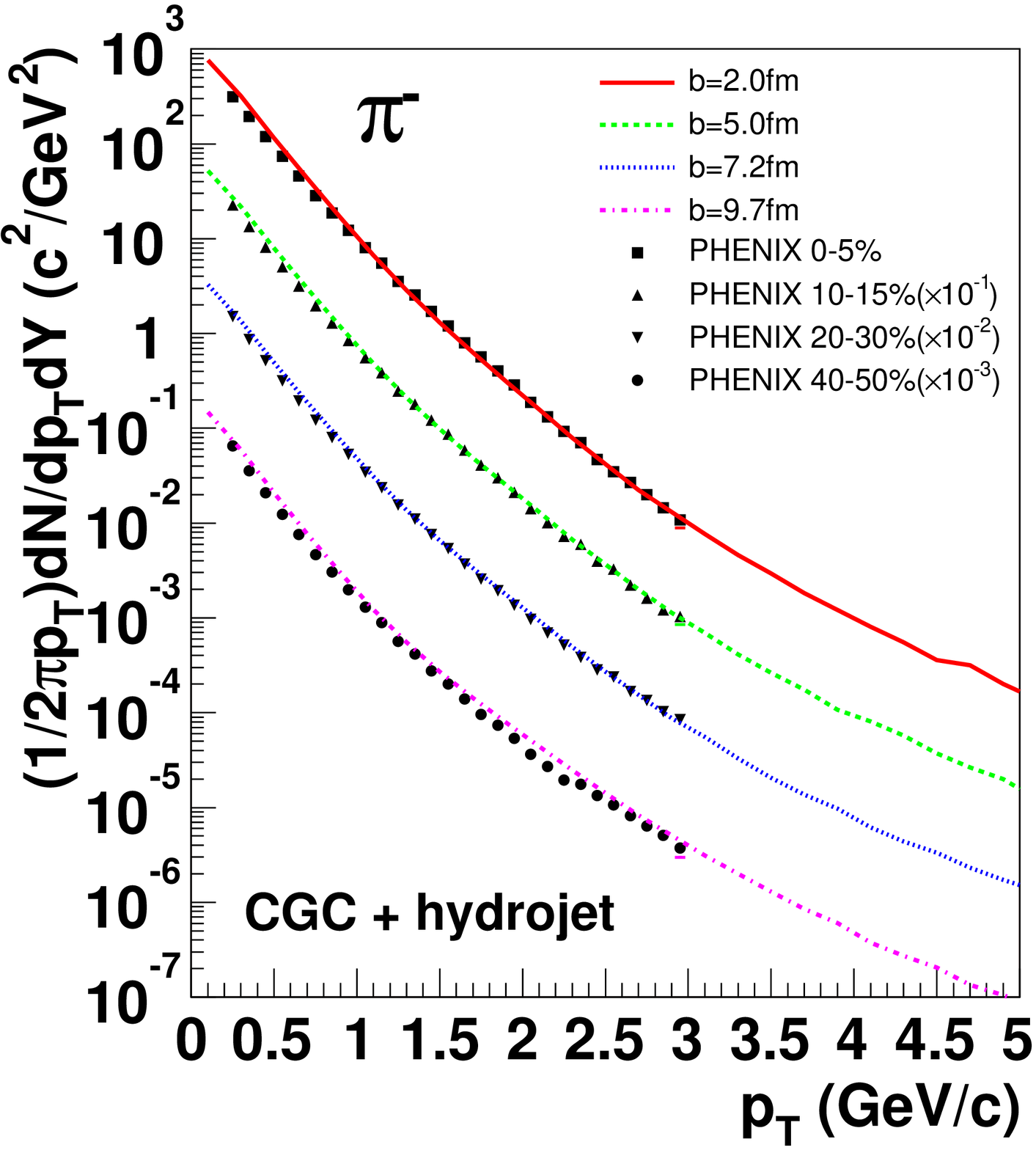}
\includegraphics[width=3.0in]{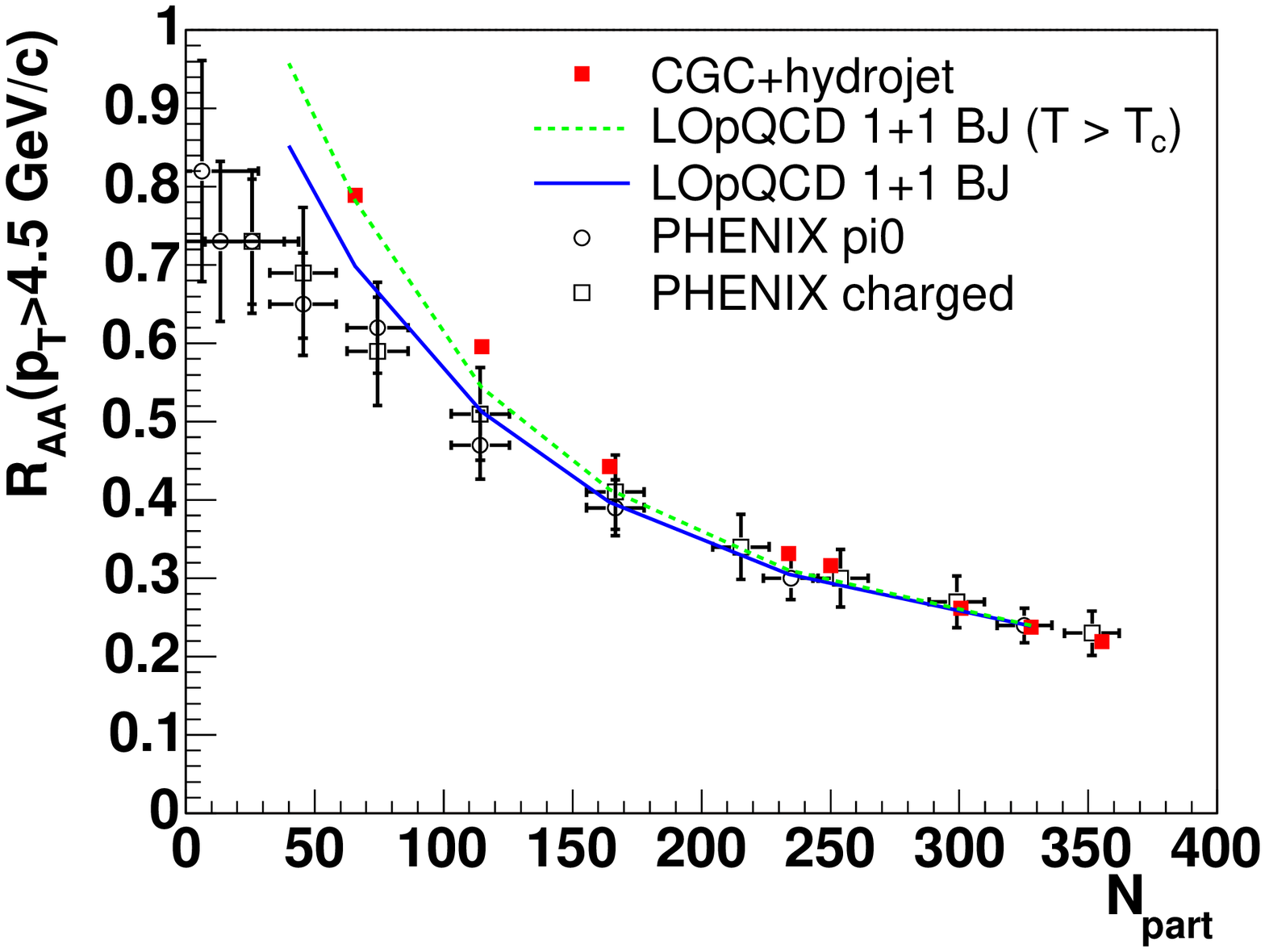}
\caption{
Transverse momentum spectra for negative pions (left)
and nuclear modification factor $R_{AA}$ (right)
 in Au + Au  collisions at $\sqrt{s_{NN}}=200$
are compared to the PHENIX data~\cite{phenix:pi}.
}
\label{fig:RAA45}
\end{figure}

We now turn to the discussion of the high $p_T$ hadron spectrum.
In our model, high $p_T$ jets suffer interaction
with the local parton density which is governed by hydrodynamic
evolution.
We only take into account parton energy loss in deconfined matter $T \ge T_c$.
In Fig.~\ref{fig:RAA45},
the result of the centrality dependence of 
the pion spectrum and
the nuclear modification factor
$R_{AA}$ integrated over $p_T>4.5$ GeV/c for $|\eta|<0.35$
are compared to PHENIX data~\cite{phenix:pi}.
Our results only account for the data up to mid-central events
and fail to reproduce data at peripheral collisions,
because neither CGC nor hydrodynamics can be applied in the
low density region.
The centrality dependence is well described by assuming
the number of participants scaling~\cite{Wang}.
However, note that our density scales as
$\rho \sim \frac{1}{\alpha_s(Q_s^2)}Q_s^2
\sim \frac{1}{\alpha_s(Q_s^2)}\rho_{\mathrm{part}}$
at mid-rapidity.
By comparison, we also plot the LOpQCD calculations with the
same initial condition
assuming (1+1)D expansion of the system ($\rho \sim 1/\tau$).
LOpQCD calculation without parton energy loss at $T<T_c = 170$ MeV
is consistent with our hydrodynamic result.
However, if there is
no restriction on the minimum temperature
in the calculation of energy loss,
we see that agreement with data becomes somewhat better.
This may indicate the contribution from hadronic interactions
at peripheral collisions.

In summary,
CGC, hydrodynamics and the energy loss of hard jets have been integrated
into one model and this dynamical approach describes the bulk properties
of RHIC data.
A systematic study within this dynamical approach is in progress.

\ack
We are grateful to
J.~Jalilian-Marian,
U.~Heinz, D.~Kharzeev, L.~McLerran, and R.~Venugopalan
for useful comments.
The work of T.H.~is supported by
RIKEN.
Research of Y.N. is supported by the DOE under Contract No.
DE-FG03-93ER40792.

\section*{References}


\end{document}